\documentclass[aps,prb]{revtex4}
\usepackage{amsmath,amssymb}
\usepackage{graphicx}
\usepackage{dcolumn}
\usepackage{bm}
\usepackage{color}
\usepackage{relsize}
\newcommand{\mbb}{\mathbb}
\newcommand{\mc}{\mathcal}

\newcommand{\tet}{\texttt}
\newcommand{\pr}{\partial}

\begin{document}
\title{Generalized WKB theory for electron tunneling in gapped $\alpha-\mc{T}_3$ lattices}
\author{Nicholas Weekes$^{1}$,
Andrii Iurov$^{1}$\footnote{E-mail contact: aiurov@mec.cuny.edu, theorist.physics@gmail.com
},
Liubov Zhemchuzhna$^{1,2}$,
Godfrey Gumbs$^{2,3}$, 
and
Danhong Huang$^{4,5}$
}

\affiliation{
$^{1}$Department of Physics and Computer Science, Medgar Evers College of City University of New York, Brooklyn, NY 11225, USA\\ 
$^{2}$Department of Physics and Astronomy, Hunter College of the City University of New York, 695 Park Avenue, New York, New York 10065, USA\\ 
$^{3}$Donostia International Physics Center (DIPC), P de Manuel Lardizabal, 4, 20018 San Sebastian, Basque Country, Spain\\ 
$^{4}$Space Vehicles Directorate, US Air Force Research Laboratory, Kirtland Air Force Base, New Mexico 87117, USA\\ 
$^{5}$Center for High Technology Materials, University of New Mexico, 1313 Goddard SE, Albuquerque, New Mexico, 87106, USA\\
}

\date{\today}

\begin{abstract}
We generalize Wentzel-Kramers-Brillouin (WKB) semi-classical equations for pseudospin-1 $\alpha-\mc{T}_3$ materials with arbitrary hopping parameter $0 < \alpha < 1$, which includes the dice lattice and graphene as two limiting cases. In conjunction with a series-expansion method in powers of Planck constant $\hbar$, we acquired and solved a system of recurrent differential equations for semi-classical electron wave functions in $\alpha-\mc{T}_3$. Making use of these obtained wave functions, we analyzed the physics-related mechanism and quantified the transmission of pseudospin-1 Dirac electrons across non-rectangular potential barriers in $\alpha-\mc{T}_3$ materials with both zero and finite band gaps. Our studies reveal several unique features, including the way in which the electron transmission depends on the energy gap, the slope of the potential barrier profile and the transverse momentum of incoming electrons. Specifically, we have found a strong dependence of the obtained transmission amplitude on the geometry-phase $\phi = \tan^{-1} \alpha$ of $\alpha-\mc{T}_3$ lattices. We believe our current findings can be applied to Dirac cone-based tunneling transistors in ultrafast analog RF devices, as well as to tunneling-current control by a potential barrier through a one-dimensional array of scatters.

\end{abstract}
\maketitle

\section{Introduction} 
\label{sec1}

The quantum states and the motion of a charged carrier in a lattice are in many ways, not much different from those described by classical dynamics. Also, the semi-classical approximation\,\cite{wkb1} is usually a useful and important tool for providing a simplified description as well as further investigations of single-particle and collective properties of electronic states with high kinetic energies in various materials.\,\cite{wkbg1, wkbg2, zalip, zalip2} Similar to standard quantum mechanics, the Wentzel-Kramers-Brillouin (WKB) approximation for an $\alpha-\mc{T}_3$ lattice can be made by expanding the electron eigenstate (or wave function) of the considered Hamiltonian as a power series of Planck's constant $\hbar$.\,\cite{zalip} Such a method is also employed for solving the second-order differential equations having coordinate-dependent coefficients, which are mathematically equivalent to the Schr{\"o}dinger equation with a spatially varying potential.\,\cite{ourS} Although most physical problems studied by this method are one-dimensional, the lately-developed WKB theory has been generalized to multiple dimensions for new two-dimensional materials.

\medskip
\par
The $\alpha-\mc{T}_3$ model\,\cite{dora1, vidal1} represents the newest, and likely, the most technologically promising class of low-dimensional materials\,\cite{thesis}  with zero-mass Dirac fermions, and it has become one of the hot spots in condensed matter physics after the discovery of graphene and its gapless, linear and relativistic low-energy band structure.\,\cite{neto} The same $\alpha-\mc{T}_3$ model bears all crucial electronic properties of graphene, yet it is still remarkably distinguished by the presence of an additional flat band at the Dirac point of its energy dispersion. This dispersionless energy band remains stable and persists in the presence of charged disorder states, external electric,  magnetic and optical fields, or a time-dependent modulation potential. Consequently, the observed energy spectrum of $\alpha-\mc{T}_3$ appears as metallic, i.e., all these  three bands intersect at the corners of Brillouin zone.

\medskip
\par
From the perspective of the atomic structure, the difference between an $\alpha-\mc{T}_3$ and graphene honeycomb lattices appears through an additional fermionic atom situated at the center of each hexagon, which is referred to as a hub C atom. The hopping integral between the hub and  one of the A and B rim atoms is different from that between nearest-neighbor rim atoms of hexagon, and the ratio of these hopping parameters is quantified by a variable $\alpha$. The maximum value for $\alpha$ is $1$ corresponding to the dice lattice, while its minimum value is $0$ for graphene for a completely decoupled set of hub atoms. From this point of view, the $\alpha-\mc{T}_3$ model can be essentially viewed as an interpolation between graphene and a dice lattice as $\alpha$ increases continuously from $0$ to $1$. Recently, models for stronger interactions with $\alpha > 1$ between the hub and rim atoms have also been proposed and explored.  

\medskip
\par
Initially designed as a purely theoretical model,\,\cite{berc} $\alpha-\mc{T}_3$, and especially the dice lattices, has recently been found in a number of existing and experimentally synthesized  materials.\,\cite{R110} These include three-layer arrangement of SrTiO$_3$/SrIrO$_3$/SrTiO$_3$ lattices\,\cite{dice1}, Lieb\,\cite{Dan2, Dan1, Dan6, Dan7} and the Kagome\,\cite{R115, R116, R117} optical lattices and waveguides\,\cite{Dan5, R133}, Josephson arrays,\,\cite{R200}, Hg$_{1-x}$Cd$_x$ quantum well\,\cite{Malc01}.  A comprehensive review of all dice-like systems with a flat band can be found in Ref.\,[\onlinecite{Add1}]. Very recently, a band structure involving  flat bands was realized in In$_{0.53}$Ga$_{0.47}$As/InP semiconductor quantum wells along with lateral geometry.\,\cite{nnfl} The $\alpha-\mc{T}_3$ model reveals a number of promising electronic,\,\cite{ourplay, Dutta1} collective,\,\cite{Malc01, ourml, fuf1} magnetic,\,\cite{piech1, piech2, tutul1, tutul2, nic1, nic2} optical,\,\cite{opt1, car1} and transport\,\cite{our20, wa20, c0} features, such as topological Dirac semimetals\,\cite{isl1, isl2} and 
tilted Dirac cone materials.\,\cite{tilt1, tilt2, tilt3}  All $\alpha-\mc{T}_3$ materials demonstrate unique symmetry and topological properties, especially in the presence of a dressing irradiation.\,\cite{dey2,dey3} For example, a photo-induced topological phase transition is observed under a non-resonant optical field\,\cite{prx1, ourPh, kisrep, kiMain, ourJAP2017ph, fnam} with a specific polarization\,\cite{dey1}, and the lattice is turned from a semimetal to Haldane-like Chern insulator in this case.

\medskip
\par
The Klein paradox, defined as complete electron tunneling independent of barrier height and width, is one of the landmark properties of all Dirac materials including graphene.\,\cite{kats}  This paradoxical behavior was predicted for  $\alpha-\mc{T}_3$ for all possible parameters $\alpha$\,\cite{alphaDice, alphaKlein, xu1, yeke} and becomes asymmetric  (i.e., observed at a finite electron incidence angle) under linearly-polarized irradiation.\,\cite{ournew, LiuM, our89} In graphene, unimpeded transmission also exists for a trapezoidal (not square) potential barrier facilitated by a finite electric field in the barrier region.\,\cite{sonin, fanwar}  The question whether such an effect could also be observed for $\alpha-\mc{T}_3$ with $\alpha > 0$ remains answered and this query becomes the major issue addressed in this paper based on the WKB approach.  Considering that the electron kinetic energy under a long-range electric potential can be very large, we believe that the WKB approximation could be the most efficient tool in solving  such a problem.

\medskip
\par
The remainder of the paper is organized as follows. In Sec.\,\ref{sec2}, we proceed with a derivation of the formalism connecting different orders of wave function expanded in powers of the Planck constant $\hbar$. Making use of this expansion, we obtain closed-form analytic expressions for the wave function including the phase factor in addition to its spatially dependent amplitude.  In Sec.\,\ref{sec3}, we apply our theory for calculating the transmission of electrons in various cases with non-square potential barriers,  such as a trapezoidal barrier imposed by a linear-potential profile. Meanwhile, we also consider gapped $\alpha-\mc{T}_3$ materials and compute their energy dispersion,  semi-classical action, classically forbidden regions, and amplitude of transmission.  Finally, some remarks are present in Sec.\,\ref{sec4} regarding derivation of the set of WKB equations and their application to electron transmission and other numerical results.

\section{Semi-Classical solution of $\alpha-\mc{T}_3$ model} 
\label{sec2}

In this section, we derive the semi-classical wave functions of a gapless $\alpha-\mc{T}_3$ lattice up to the first  order of the series expansion  in powers of $\hbar$. An important step of this derivation involves calculating the semi-classical action of which the spatial derivative in Eq.\,\eqref{sp} corresponds to a position-dependent longitudinal electron momentum given in Eq.\,\eqref{sp2}.  The mere knowledge of the action $S(\xi)$, or momentum component $\pi_x(\xi)$, in terms of the scaled length $\xi$ defined in Eq.\ (\ref{units}), only provides a possibility to evaluate the electron transmission through a non-square barrier under a finite  longitudinal electric field. However, an analytical expression for wave function either does not exist or is too complicated to obtain   as we deal with gapped $\alpha-\mc{T}_3$ materials.  

\medskip
\par
We begin our study with the pseudospin-1 low-energy Hamiltonian for the $\alpha-\mc{T}_3$ model, i.e.,  

\begin{equation}
\label{H0}
\hat{\mc{H}}_{\alpha} (\mbox{\boldmath$k$} \, \vert \, \tau, \phi) = \hbar v_F 
\left[
\begin{array}{ccc}
0 & k^\tau_- \, \cos \phi &  0 \\
k^\tau_+ \, \cos \phi & 0 & k^\tau_- \, \sin \phi \\
0 & k^\tau_+ \, \sin \phi & 0
\end{array}
\right] \, ,  
\end{equation}
where $k^\tau_\pm = \tau k_x \pm i k_y$, and valley index $\tau = \pm 1$ distinguishes electronic states associated with $K$ and $K'$ valleys.  Phase $\phi$ $(0\leq\phi\leq\pi/4)$ introduced in Eq.\,\eqref{H0} relates to $\alpha$ parameter by $\alpha = \tan \phi$, and therefore, the limiting cases for graphene and dice lattice correspond to  $\phi=0$ and $\phi=\pi/4$, respectively.   
  
\medskip
\par
The Hamiltonian in Eq.\,\eqref{H0} could be constructed using the following two $\phi$-dependent $3 \times 3$ matrices $\hat{\mbox{\boldmath$S$}}(\phi)= \left\{  \hat{S}_x(\phi),\,\hat{S}_y(\phi) \right\}$, where

\begin{equation}
\label{Sxp}
\hat{S}_x(\phi) = \left[
\begin{array}{ccc}
 0 & \cos \phi & 0 \\
 \cos \phi & 0 & \sin \phi \\
 0 & \sin \phi & 0
\end{array}
\right] \ ,
\end{equation}

\begin{equation}
\label{Syp}
\hat{S}_y(\phi) = i \,\left[
\begin{array}{ccc}
 0 & -\cos \phi & 0 \\
 \cos \phi & 0 & -\sin \phi \\
 0 & \sin \phi & 0
\end{array}
\right] \ .
\end{equation}
As a result, we have

\begin{equation}
\label{HamG}
\hat{\mc{H}}_{\alpha} (\mbox{\boldmath$k$} \, \vert \, \tau, \phi)  = v_F \, \hat{\mbox{\boldmath$S$}}(\phi) \cdot \left\{ -i \hbar\mbox{\boldmath$\nabla$}_\tau \right\} + V(x) \, \hat{\Sigma}^{(3)}_{0} \ ,
\end{equation}
where $\hat{\Sigma}^{(3)}_{0}$ is a $3\times 3$ unit matrix, $V(x)$ represents a position-dependent electrostatic potential, and $\mbox{\boldmath$\nabla$}_\tau=\{ \tau \pr/\pr x,\,\pr/\pr y \}$. In fact, the introduced matrices presented in Eqs.\,\eqref{Sxp} and  \eqref{Syp} are a $\phi$-dependent generalization of $3 \times 3$ Pauli matrices written as

\begin{equation}
\hat{\Sigma}^{(3)}_{x} = \frac{1}{\sqrt{2}} \, \left[
\begin{array}{ccc}
 0 & 1 & 0 \\
 1 & 0 & 1 \\
 0 & 1 & 0
\end{array}
\right] \ ,
\label{sig1}
\end{equation}

\begin{equation}
\hat{\Sigma}^{(3)}_{y} = \frac{i}{\sqrt{2}} \, \left[
\begin{array}{ccc}
 0 & -1 & 0 \\
 1 & 0 & -1 \\
 0 & 1 & 0
\end{array}
\right] \ .
\label{sig2}
\end{equation}
where $\phi = \pi/4$ is taken. On the other hand, 
for $\phi \rightarrow 0$, matrices in Eqs.\,\eqref{Sxp} and \eqref{Syp} reduce to $2 \times 2$ spin-$1/2$ Pauli matrices used to define a Dirac Hamiltonian in graphene. 
One sometimes employs the third Pauli matrix, defined as  

\begin{equation}
\label{szdice}
\hat{\Sigma}^{(3)}_{z} = \left[
\begin{array}{ccc}
 1 & 0 & 0 \\
 0 & 0 & 0 \\
 0 & 0 & -1
\end{array}
\right] \ , 
\end{equation}
so as to introduce an energy gap to a pseudospin-$1$ Hamiltonian.\,\cite{Gusgap}
\medskip

Three eigen-energies associated with the Hamiltonian in Eq.\,\eqref{H0} are easily found to be

\begin{equation}
\label{vc}
\varepsilon^{\gamma=\pm 1}_{\tau, \, \phi}(\mbox{\boldmath$k$}) =  \gamma \hbar v_F k
\end{equation}
with $\gamma = - 1$ ($\gamma = + 1$) for the valance (conduction) band, and 

\begin{equation}
\label{fb}
\varepsilon^{\gamma=0}_{\tau, \, \phi}(\mbox{\boldmath$k$}) = 0\ .
\end{equation}
for the remaining flat (or dispersionless) band. 
Here, all three bands in Eqs.\,\eqref{vc} and \eqref{fb} do not show any dependence on phase $\phi$ (or parameter $\alpha$).  
Furthermore, two wave functions corresponding to the valence and conduction bands in Eq.\,\eqref{vc} take the form 

\begin{equation}
\label{Eig1}
\Psi^{\gamma=\pm 1}_{\tau, \, \phi}(\mbox{\boldmath$k$})  = \frac{1}{\sqrt{2}} \left[
\begin{array}{c}
\tau \cos \phi \,\, \tet{e}^{- i \tau \theta_{ \bf k}}  \\
\gamma \\
\tau \sin \phi \,\, \tet{e}^{+ i \tau \theta_{ \bf k}} 
\end{array}
\right]\ ,
\end{equation}
where $\theta_{\bf k} = \arctan (k_y/k_x)$ is the angle of wave vector $\mbox{\boldmath$k$} = \{k_x, k_y\}$ made with the $x$-axis. The other 
wave function for the flat band is 

\begin{equation}
\label{Eig2}
\Psi^{\gamma=0}_{\tau, \, \phi}(\mbox{\boldmath$k$}) = \left[
\begin{array}{c}
\sin \phi \,\, \tet{e}^{- i \tau \theta_{\bf k}}  \\
0 \\
- \cos \phi \,\, \tet{e}^{+ i \tau \theta_{\bf k}} 
\end{array}
\right]\ . 
\end{equation}
Here, we would like to indicate that the energy bands in Eqs.\,\eqref{vc} and \eqref{fb}, as well as the wave functions in Eqs.\,\eqref{Eig1} and \eqref{Eig2},  
are obtained for a spatially-uniform potential independent of position coordinates $x$ and $y$.
\medskip

As a generalization, we now consider an $x$-dependent potential $V(x)$ so that the translational symmetry is kept only along the $y$ direction, 
and the wave function changes to $\Psi(x,y) \backsim \psi(x)\,\tet{e}^{i k_y \, y}$. Correspondingly, the previous Hamiltonian in Eq.\,\eqref{HamG} is modified into 

\begin{equation}
\label{H02}
\hat{\mc{H}} (x, k_y \, \vert \, \tau, \phi) = \hbar v_F
\left[
\begin{array}{ccc}
V(x) & \cos \phi\,(- i \hbar \, \tau \, \pr/\pr x - i p_y) &  0 \\
\cos \phi\,(- i \hbar \, \tau \, \pr/\pr x + i p_y) & V(x) & \sin \phi\,(- i \hbar \, \tau \, \pr/\pr x - i p_y)\\
0 & \sin \phi\,(- i \hbar \, \tau \, \pr/\pr x - i p_y) & V(x)
\end{array}
\right] \, ,
\end{equation}
where $p_y = \hbar k_y$ is conserved in the tunneling process.
Following the approach and notations adopted in Ref.\,[\onlinecite{zalip}], we rewrite our Hamiltonian in Eq.\,\eqref{H02} and the corresponding eigenvalue equation 
through the following dimensionless variables, i.e., $x \rightarrow \xi$, $E \rightarrow \varepsilon$, $p_{x,y} \rightarrow \pi_{x,y}$ and 

\begin{equation}
\xi = \frac{x}{W_B}\ ,\ \ \ \ \ \ \varepsilon = \frac{E}{V_0} \,\,\, \text{\ \ and\ \ } \,\,\, \nu(x) = \frac{V(x)}{V_0} \ ,\ \ \ \ \ \ \pi_y = \frac{v_y p_y}{V_0} \ .
\label{units}
\end{equation} 
Finally, using the fact that $\pr/\pr x \rightarrow \pr / (W_B \, \pr \xi)$, we replace Planck constant $\hbar$ by a dimensionless one $\hbar_0$, yielding

\begin{equation}
\hbar_0 \leftrightarrow \left(\frac{v_F}{W_BV_0}\right)\ \hbar \ ,
\end{equation}
where $V_0$ is the height of the square barrier in the absence of a longitudinal electric field.   In Eq.\,\eqref{units}, we avoided using an energy scale $E_F = \hbar v_Fk_F \backsim \hbar v_F \, 1/L_0$ with a unit length $L_0$ for electron Fermi energy  since we do not  wish to introduce additional $\hbar$-related terms in the eigenvalue equation. This implies that the energy scale for incoming particles can be large classically and not limited by values in units of $E_F$. 

\medskip 
\par
By using the dimensionless variables defined in Eq.\,\eqref{units}, the eigenvalue equation  becomes 

\begin{eqnarray}
\label{new1}
&&\hat{\mc{H}}_{\alpha} (\xi, \pi_y \, \vert \, \tau, \phi) \, \Psi^{\,\gamma}(\xi, \pi_y \, \vert \, \phi, \tau) = \varepsilon \, \Psi^{\,\gamma}(\xi, \pi_y \, \vert \, \phi, \tau)\ ,\\
&&\Psi^{\,\gamma}(\xi, \pi_y \, \vert \, \phi, \tau) = \Psi_x ^{\,\gamma}(\xi \, \vert \, \phi, \tau) \, \tet{exp}\left(i \frac{\pi_y\eta}{\hbar_0} \right) =  \left[
\begin{array}{c}
\phi_A (\xi \, \vert \, \phi, \tau) \\
\phi_H (\xi) \\
\phi_B (\xi \, \vert \, \phi, \tau)
\end{array}
\right]
\tet{exp}\left( i \frac{\pi_y\eta}{\hbar_0} \right)\ , 
\label{new2}
\end{eqnarray}
and the Hamiltonian in Eq.\,\eqref{new1} is now  

\begin{eqnarray}
\nonumber
\hat{\mc{H}}_{\alpha} (\xi, \pi_y \, \vert \, \tau, \phi) &=& 
\hat{\Sigma}^{(3)}_{0} \, \nu(x) + \hat{\Sigma}^{(3)}_{x}(\phi) \, \left(
- i \hbar_0 \, \tau \, \frac{\pr}{\pr \xi }
\right) + \hat{\Sigma}^{(3)}_{y}(\phi)  \, \pi_y \\
&=& \left[
\begin{array}{ccc}
\nu(x) &  \cos\phi\left( - i \hbar_0\tau \, \pr / \pr \xi - i \pi_y \right) & 0 \\
\cos\phi\left( - i \hbar_0\tau \, \pr/\pr \xi + i \pi_y \right) &  \nu(x)
&  \sin\phi\left( - i \hbar_0\tau \, \pr/\pr \xi - i \pi_y \right)\\
0 & \sin\phi\left( - i \hbar_0\tau \, \pr/\pr \xi + i \pi_y \right) & \nu(x)
\end{array}
\right]\ ,
\label{h_new} 
\end{eqnarray}
where $\hat{\Sigma}^{(3)}_{0}$ represents a $3\times 3$ unit matrix.
\medskip 

We know that the standard WKB approach is based on a series expansion of the sought wave function over $\hbar_0 \backsim \hbar$, namely, 

\begin{equation}
\label{expand}
\Psi(\xi, \pi_y \, \vert \, \phi, \tau)  = \tet{exp} \left\{\frac{i}{\hbar_0} \, \mbb{S}(x) \right\}\, \sum \limits_{\lambda = 0}^{\infty} (- i \hbar_0)^\lambda \, \Psi_\lambda (x) = \tet{exp} \left\{ \frac{i}{\hbar_0} \, \mbb{S}(x) \right\}\, \left[\Psi_0 (x) - i \hbar_0 \, \Psi_1 (x) - \hbar_0^2 \, \Psi_2 (x)  + \cdots \, \right]\ ,
\end{equation}
where $\mbb{S}(x)$ represents the semi-classical action in the WKB approximation,  and our goal is obtaining a differential equation with respect to $x$, which connects consecutive  terms in the  expansion in Eq.\,\eqref{expand}. From Eq.\,\eqref{h_new}, however, we find that only a term involving $\hbar_0$ can serve for this purpose. Explicitly, we write down such an equation as

\begin{equation}
\label{transeq}
\hat{\Sigma}^{(3)}_{x} (\phi) \, \left\{\, \frac{\pr}{\pr \xi} \,  \Psi_{\lambda}(\xi, \pi_y \, \vert \, \phi, \tau)  \, \right\}
- \frac{1}{\sqrt{2}} \, \hat{\mbb{O}}_{\,T} (\xi, \pi_y \, \vert \, \phi \, \tau) \, \Psi_{\lambda+1} (\xi, \pi_y \, \vert \, \phi, \tau) 
= 0 \ ,
\end{equation}
where $\lambda = 0, \,1, \,2, \, 3, \,\cdots$, and $\Psi_{\lambda =  - 1}(\xi, \pi_y \, \vert \, \phi, \tau)\equiv 0$. 
Here, the transport operator $\hat{\mbb{O}}_{\,T} (\xi, \pi_y \, \vert \, \phi, \, \tau)$, connecting consequent terms of expansion in Eq.\,\eqref{expand}, is easily found to be

\begin{equation}
\label{to1}
\hat{\mbb{O}}_{\,T} (\xi, \pi_y \, \vert \, \phi, \tau) = 
\left[
\begin{array}{ccc}
\kappa(\xi) & \cos \phi \,\left(\tau\pr\mbb{S}(\xi)/\partial\xi- i \pi_y \right) & 0 \\
\cos \phi \,\left(\tau\pr\mbb{S}(\xi)/\partial\xi + i \pi_y \right) & \kappa(\xi)
& \sin \phi \,\left(\tau\pr\mbb{S}(\xi)/\partial\xi - i \pi_y \right) \\
0 & \sin \phi \,\left(\tau\pr\mbb{S}(\xi)/\partial\xi + i \pi_y \right) & \kappa(\xi)
\end{array}
\right]\ , 
\end{equation}
where $\kappa(\xi) = \nu(\xi) - \varepsilon$. Specifically, by setting $\lambda = - 1$, Eq.\,\eqref{transeq} gives rise to 

\begin{equation}
\label{zeroord}
\hat{\mbb{O}}_{\,T} (\xi, \pi_y \, \vert \, \phi, \tau) \, \Psi_0 (\xi, \pi_y \, \vert \, \phi, \, \tau) = 0 \ .
\end{equation}
For a linear homogeneous Eq.\,\eqref{zeroord}, a non-trivial solution exists only if its determinant is zero, i.e.,  

\begin{equation}
\left(\varepsilon - \nu(\xi) \right)\, \left[ \, \left(\frac{\pr \, \mbb{S}(\xi)}{\pr\xi} \right)^2 + \pi_y^2 - \left(\varepsilon - \nu(\xi)\right)^2 \, \right] = 0 \ ,
\label{new-4}
\end{equation}
which is independent of $\phi$.
Generally speaking, we know $\nu(\xi) \neq \varepsilon$, and then Eq.\,\eqref{new-4} leads us to

\begin{equation}
\label{sp}
\mbb{S}(\xi) -\mbb{S}(\xi_0) = \int\limits_{\xi_0}^{\xi} \pi_x(\eta) \, d \eta \ , 
\end{equation}
where 

\begin{equation}
\label{sp2}
\pi_x(\xi) = \pm \sqrt{ \left[  \varepsilon - \nu(\xi) \right]^2 - \pi_y^2} 
\end{equation}
represents the position-dependent longitudinal momentum of electrons, while the transverse momentum $\pi_y$ remains as a constant in the tunneling process. 
\medskip 

As a next step, we want to find the leading-order wave function $\Psi_1(x)$. Although Eq.\,\eqref{zeroord} appears as an eigenvalue problem, it is actually much more 
complicated since $\mbb{S}(\xi)$ in Eq.\,\eqref{sp} and $\pi_x(\xi)$ in Eq.\,\eqref{sp2} also depend on particle energy $\varepsilon$. 
In fact, Eq.\,\eqref{zeroord} could be utilized to find various components of the following zero-order wave function 

\begin{equation}
\label{zo}
\Psi_0 (\xi,\pi_y \, \vert \, \phi, \tau) = \left[
\begin{array}{c}
\varphi_A^{(0)} (\xi \, \vert \, \phi, \tau) \\
\varphi_H^{(0)} (\xi) \\
\varphi_B^{(0)} (\xi \, \vert \, \phi, \tau)
\end{array}
\right]\ ,
\end{equation}
such that

\begin{eqnarray}
\kappa(\xi)\,\varphi_A^{(0)}(\xi)+\cos\phi\,\left[\tau\pi_x(\xi)-i\pi_y\right]\,\varphi_H^{(0)}(\xi)&=&0\ ,\\
\sin\phi\,\left[\tau\pi_x(\xi)+i\pi_y\right]\,\varphi_H^{(0)}(\xi)+\kappa(\xi)\,\varphi_B^{(0)}(\xi)&=&0\ .
\end{eqnarray}
Consequently, the wave function in Eq.\,\eqref{zo} can be rewritten as

\begin{equation}
\label{wf0theta}
\Psi_0 (\xi \, \vert \, \phi, \, \tau) = \left[
\begin{array}{c}
\cos \phi \,\, \Theta(\xi \, \vert \, \tau) \\
-1 \\
\sin \phi \,\, \Theta^{\star}(\xi \, \vert \, \tau)
\end{array}
\right]\,\varphi_H^{(0)} (\xi) \ ,
\end{equation}
where $\Theta^{\star}(\xi \, \vert \, \tau)$ represents the complex conjugate of $\Theta(\xi \, \vert \, \tau)$, which is given by 

\begin{equation}
\label{theta}
\Theta(\xi \, \vert \, \tau) = \frac{1}{\kappa(\xi)} \, \left[ \tau\pi_x(\xi) - i \pi_y \right] = 
- \tau \, \tet{exp}\left[ - i\tau \, \theta_{\bf k}(\xi) \right]\, .
\end{equation}
Here, $\theta_{\bf k}(\xi) = \tan^{-1} [k_x(\xi)/k_y]$ is the angle of wave vector $\mbox{\boldmath$k$}= \{k_x(\xi), k_y\} = 1/\hbar \, \{\pi_x(\xi), \pi_y \}$ with respect to the $x$-axis.  It is clear from Eq.\,\eqref{theta} that $\Theta(\xi \, \vert \, \tau)$ does depend on the valley index $\tau$ and position $\xi$ but not on $\phi$. Moreover, the spatial dependence of $\varphi_H^{(0)} (\xi)$ in Eq.\,\eqref{wf0theta} still needs to be determined.

\medskip
\par
Moreover, by taking $\lambda = 0$, we get from the recurrence equation in Eq.\,\eqref{transeq} that 

\begin{equation}
\label{transeq1}
\hat{\mbb{O}}_{\,T} (\xi, \pi_y \, \vert \, \phi, \, \tau) \, \Psi_{1} (\xi, \pi_y \, \vert \, \phi, \tau) = 
\sqrt{2} \, \hat{\Sigma}^{(3)}_{x} (\phi)  \, \frac{\pr}{\pr \xi} \,  \Psi_{0}(\xi, \pi_y \, \vert \, \phi, \tau)\ ,
\end{equation} 
where $\Psi_{0}(\xi, \pi_y \, \vert \, \phi, \tau)$ has already been obtained expect for the spatial dependence on $\varphi_H^{(0)} (\xi)$. However, 
we still do not know the exact form of $\Psi_{1} (\xi, \pi_y \, \vert \, \phi, \tau)$. 
Mathematically, $\Psi_{1} (\xi, \pi_y \, \vert \, \phi, \tau)$ can be constructed from a linear combination of three arbitrary orthogonal state vectors 
$ \big \vert \mbox{\boldmath$v$}_1 \bigr>$, $ \big \vert \mbox{\boldmath$v$}_2 \bigr>$ and $\big \vert \mbox{\boldmath$v$}_3 \bigr>$
in a three-dimensional spinor space. 
Let us first choose $ \big \vert \mbox{\boldmath$v$}_1 \bigr>$ to be the spinor part of $\Psi_0 (\xi \, \vert \, \phi,  \tau)$ in Eq.\,\eqref{wf0theta}, yielding  

\begin{equation}
\big \vert \mbox{\boldmath$v$}_1(\xi \, \vert \, \phi,  \tau) \bigr>  = \left[
\begin{array}{c}
\cos \phi \,\, \Theta(\xi \, \vert \, \tau) \\
-1 \\
\sin \phi \,\, \Theta^{\star}(\xi \, \vert \, \tau)
\end{array}
\right]\ .
\label{vec1}
\end{equation}
With the given spinor state $\big \vert \mbox{\boldmath$v$}_1(\xi\,\vert\,\phi,\tau) \bigr>$ in Eq.\,\eqref{vec1}, we can choose freely the remaining 
$\big \vert \mbox{\boldmath$v$}_2(\xi\,\vert\,\phi,\tau) \bigr>$  and $\big \vert \mbox{\boldmath$v$}_3(\xi\,\vert\,\phi,\tau) \bigr>$ as long as all of them are mutually orthogonal to each other.  By referencing wave functions in Eqs.\,\eqref{Eig1} and \eqref{Eig2} for incident particles, we take accordingly 

\begin{eqnarray}
\big \vert \mbox{\boldmath$v$}_2(\xi \, \vert \, \phi,  \tau) \bigr>  &=& \left[
\begin{array}{c}
\cos \phi \,\, \Theta(\xi \, \vert \, \tau) \\
+1 \\
\sin \phi \,\, \Theta^{\star}(\xi \, \vert \, \tau)
\end{array}
\right]\ ,\\
\big \vert \mbox{\boldmath$v$}_3(\xi \, \vert \, \phi,  \tau) \bigr> &=& \left[
\begin{array}{c}
\sin \phi \,\, \Theta(\xi \, \vert \, \tau) \\
0 \\
- \cos \phi \,\, \Theta^{\star}(\xi \, \vert \, \tau)
\end{array}
\right] \ . 
\end{eqnarray}
Consequently, $\Psi_{1} (\xi, \pi_y \, \vert \, \phi, \tau)$ can be formally written as 

\begin{equation}
\label{3exp}
\Psi_{1} (\xi, \pi_y \, \vert \, \phi, \tau) = \varphi_H^{(0)} (\xi)\, \big \vert\mbox{\boldmath$v$}_1(\xi\,\vert\,\phi,\tau)\bigr> + 
\varphi_H^{(1,2)} (\xi)\,\big \vert\mbox{\boldmath$v$}_2(\xi\,\vert\,\phi,\tau)\bigr>+\varphi_H^{(1,3)} (\xi)\,\big \vert\mbox{\boldmath$v$}_3(\xi\,\vert\,\phi,\tau)\bigr>\ . 
\end{equation}
\medskip

Now, substituting Eq.\,\eqref{3exp} into Eq.\,\eqref{transeq1}, we find

\begin{eqnarray}
\nonumber
&&\bigr< \mbox{\boldmath$v$}_1(\xi\,\vert\,\phi,\tau)\,\big \vert \, \hat{\mbb{O}}_{\,T} (\xi, \pi_y \, \vert \, \phi, \tau)\,\Psi_{1} (\xi, \pi_y \, \vert \, \phi, \tau)\bigr>\\
&=&\bigr< \mbox{\boldmath$v$}_1(\xi\,\vert\,\phi,\tau)\,\big \vert \, \hat{\mbb{O}}_{\,T} (\xi, \pi_y \, \vert \, \phi,\tau)\,\big\vert \left\{\big\vert\,
\varphi_H^{(0)} (\xi)\, \big \vert\mbox{\boldmath$v$}_1(\xi\,\vert\,\phi,\tau)\bigr> + 
\varphi_H^{(1,2)} (\xi)\,\big \vert\mbox{\boldmath$v$}_2(\xi\,\vert\,\phi,\tau)\bigr>+\varphi_H^{(1,3)} (\xi)\,\big \vert\mbox{\boldmath$v$}_3(\xi\,\vert\,\phi,\tau)\bigr>\right\}\bigr> 
= 0\ \ \ \ \ \ \ \
\end{eqnarray}
\medskip
since $\big \vert \mbox{\boldmath$v$}_j(\xi\,\vert\,\phi,\tau) \bigr>$ for $j=1,\,2,\,3$ are orthogonal to each other and 
$\hat{\mbb{O}}_{\,T} (\xi, \pi_y \, \vert \, \phi, \tau)\,\big \vert\, \mbox{\boldmath$v$}_1(\xi\,\vert\,\phi,\tau) \bigr>=0$ from Eq.\,\eqref{new-4}. 
Consequently, using Eq.\,\eqref{transeq1} we get 

\begin{equation}
\label{sig}
\bigr< \mbox{\boldmath$v$}_1(\xi\,\vert\,\phi,\tau)\, \big \vert \,  \hat{\Sigma}^{(3)}_{x} (\phi)  \, \frac{\pr}{\pr \xi} \,  \Psi_{0}(\xi, \pi_y \, \vert \, \phi, \tau)\,\bigr > = 
\bigr< \mbox{\boldmath$v$}_1(\xi\,\vert\,\phi,\tau)\, \big \vert \,  \hat{\Sigma}^{(3)}_{x} (\phi)  \, \frac{\pr}{\pr \xi} \, \big \vert \,\bigr< \mbox{\boldmath$v$}_1(\xi\,\vert\,\phi,\tau)\,
\bigr>= 0\ .
\end{equation}
Here, Eq.\,\eqref{sig} can be written as

\begin{eqnarray}
\label{mainEq}
&& \left[ 
\Theta(\xi \, \vert \,\tau) + \Theta^{\star}(\xi \, \vert \,\tau)
\right] 
\, \frac{\pr \, \varphi_H^{(0)} (\xi)}{\pr \xi} \\
\nonumber 
&& + \frac{1}{2} \, 
\left\{
\,
\left[
\frac{\pr \, \Theta(\xi \, \vert \,\tau)}{\pr \xi} + \frac{\pr \, \Theta^{\star}(\xi \, \vert \,\tau)}{\pr \xi} \,
\right] + 
\left[
\frac{\pr \, \Theta(\xi \, \vert \,\tau)}{\pr \xi} - \frac{\pr \, \Theta^{\star}(\xi \, \vert \,\tau)}{\pr \xi} \,
\right] \, \cos2 \phi \,
\right\} \, \varphi_H^{(0)} (\xi) = 0 \ .
\end{eqnarray}
Writing $\Gamma_{(\pm)}(\xi \, \vert \,\tau) = \Theta(\xi \, \vert \,\tau) \pm  \Theta^{\star}(\xi \, \vert \,\tau)$, we finally arrive at the equation:   

\begin{equation}
\label{mainEq2}
\Gamma_{(+)}(\xi \, \vert \,\tau) \, \frac{\pr \, \varphi_H^{(0)} (\xi)}{\pr \xi} + \frac{1}{2} \, \left[
\frac{\pr}{\pr \xi} \, \Gamma_{(+)}(\xi \, \vert \,\tau) + 
\cos2 \phi\,\frac{\pr}{\pr \xi} \, \Gamma_{(-)}(\xi \, \vert \,\tau)
\right] \, \varphi_H^{(0)} (\xi) = 0 \ ,
\end{equation}
in which we have

\begin{equation}
\mc{F}(\xi \, \vert \, \phi, \tau) \equiv\frac{\pr}{\pr \xi} \left[
\Gamma_{(+)}(\xi \, \vert \,\tau) + 
\cos2 \phi \, \Gamma_{(-)}(\xi \, \vert \,\tau) 
\right] = \frac{ 2 i \, \pi_y}{\pi\xi^{3/2}} \,\left[  
\cos2 \phi \, \pi_x(\xi) - i \pi_y
\right] \, \frac{d\pi_x(\xi)}{d \xi}\ ,
\end{equation}
which is obtained directly from Eq.\,\eqref{theta}. 
\medskip

\begin{figure} 
\centering
\includegraphics[width=0.65\textwidth]{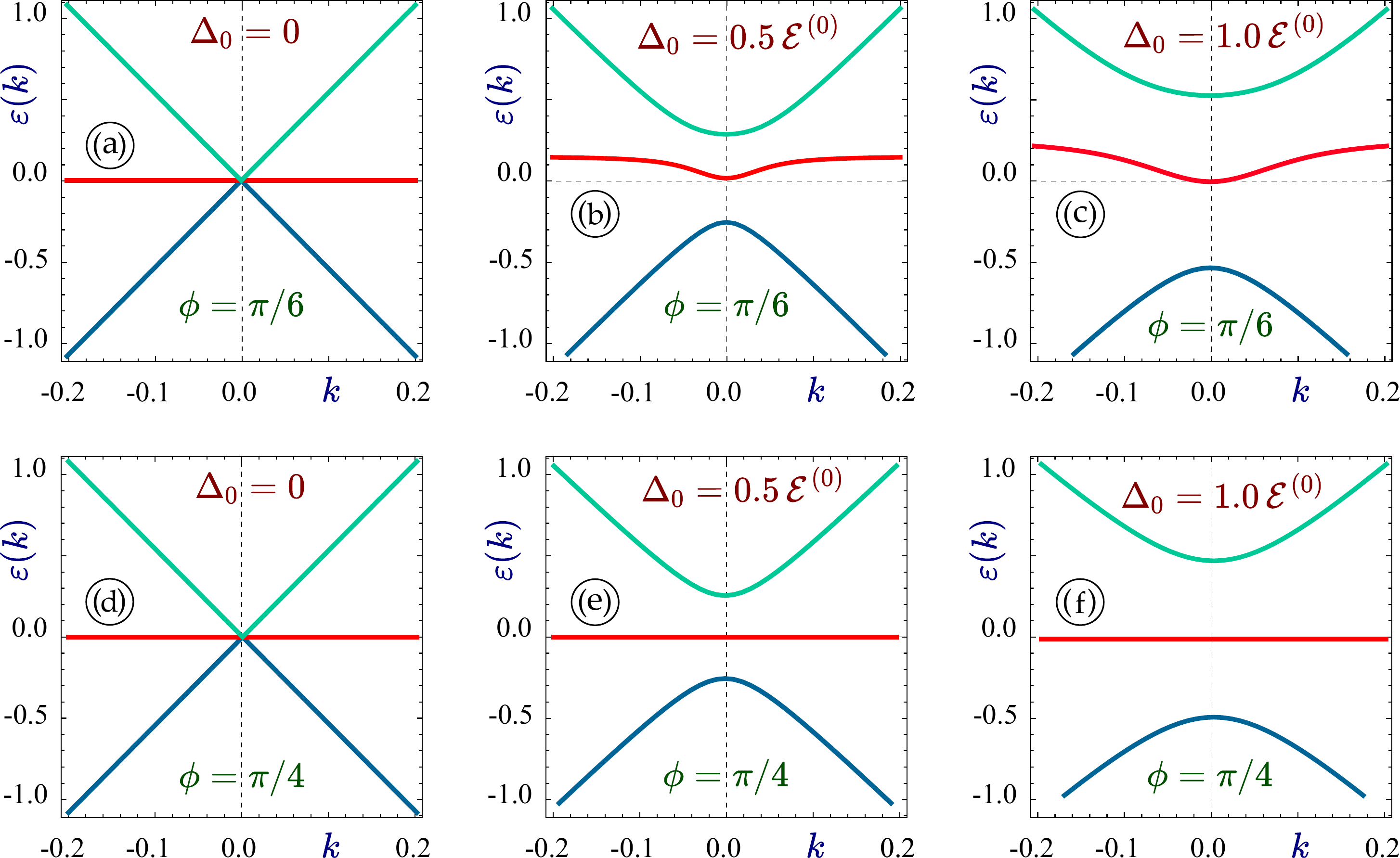}
\caption{(Color online) Calculated energy dispersion relations (shown shaded)  $\varepsilon(k,\Delta_0 \, \vert \, \phi)$ in $\alpha-\mc{T}_3$ based on Eq.\,\eqref{ed1}. Panels $(a)$--$(c)$ correspond to phase $\phi = \pi/6$, while panels $(d)$--$(f)$ to a dice lattice with $\phi = \pi/4$. The left column  relates to the case with $\Delta_0/{\cal E}^{(0)}=0$; the middle column to $\Delta_0/{\cal E}^{(0)}=0.5$; and the right column to $\Delta_0/{\cal E}^{(0)}=1.0$.  Here, the Fermi wave number $k_F$ is taken as an unit for $k$, and ${\cal E}^{(0)}=\hbar v_Fk_F$ as an unit for energy with Fermi velocity $v_F$.}
\label{FIG:1}
\end{figure}

\begin{figure} 
\centering
\includegraphics[width=0.65\textwidth]{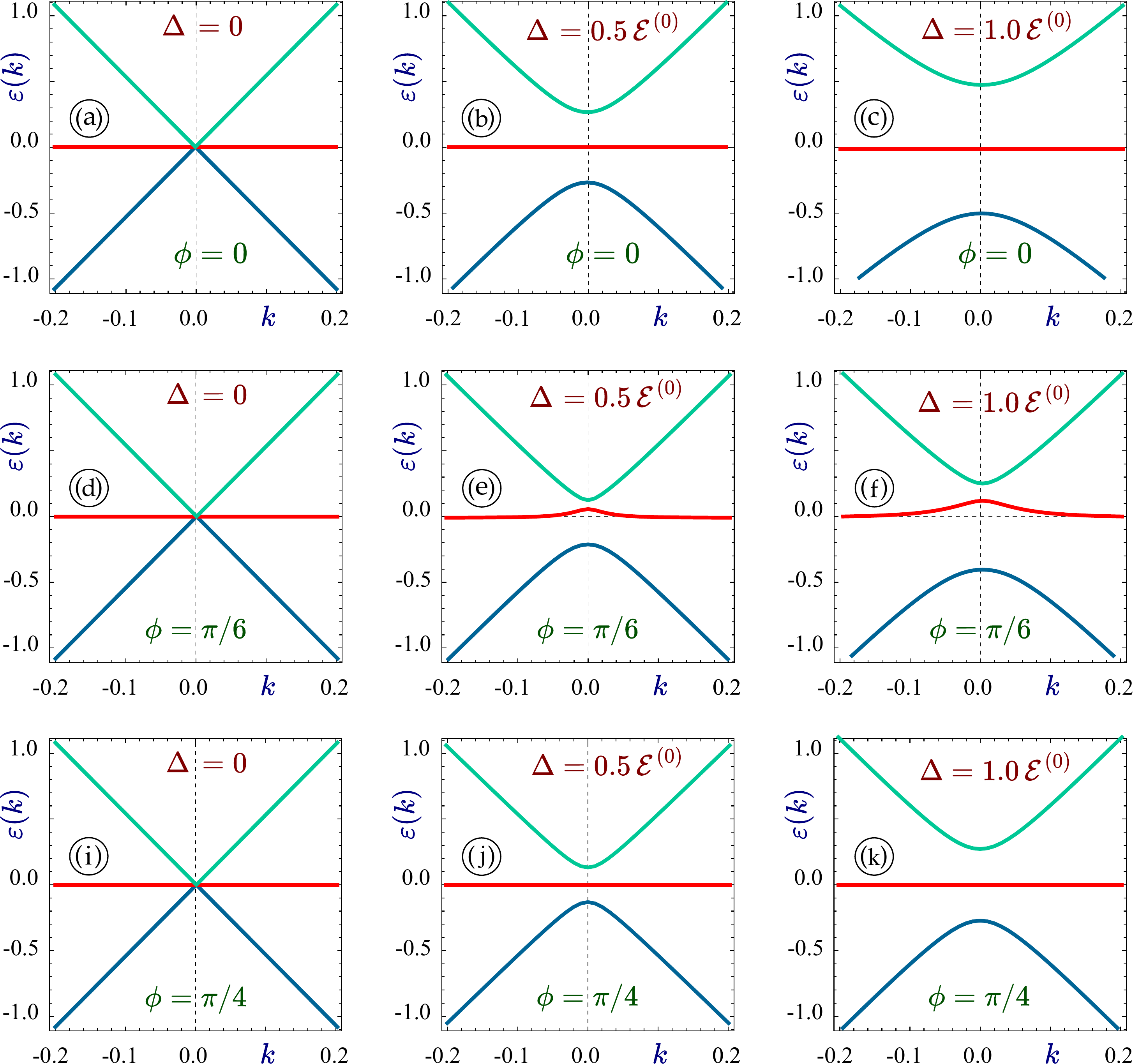} 
\caption{(Color online) Calculated energy dispersion relations $\varepsilon(k,\Delta\, \vert \, \phi)$ in $\alpha-\mc{T}_3$ based on Eq.\,\eqref{ed2} in the presence of a finite energy gap $\Delta$. Panels $(a)$--$(c)$ correspond to phase $\phi = 0$, panels $(d)$--$(f)$ to $\phi = \pi/6$, and panels $(i)$--$(k)$ to $\phi = \pi/4$. The left column  relates to the case with $\Delta/{\cal E}^{(0)}=0$; the middle column to $\Delta/{\cal E}^{(0)}=0.5$; 
and the right column to $\Delta/{\cal E}^{(0)}=1.0$.}
\label{FIG:2}
\end{figure}

\begin{figure} 
\centering
\includegraphics[width=0.65\textwidth]{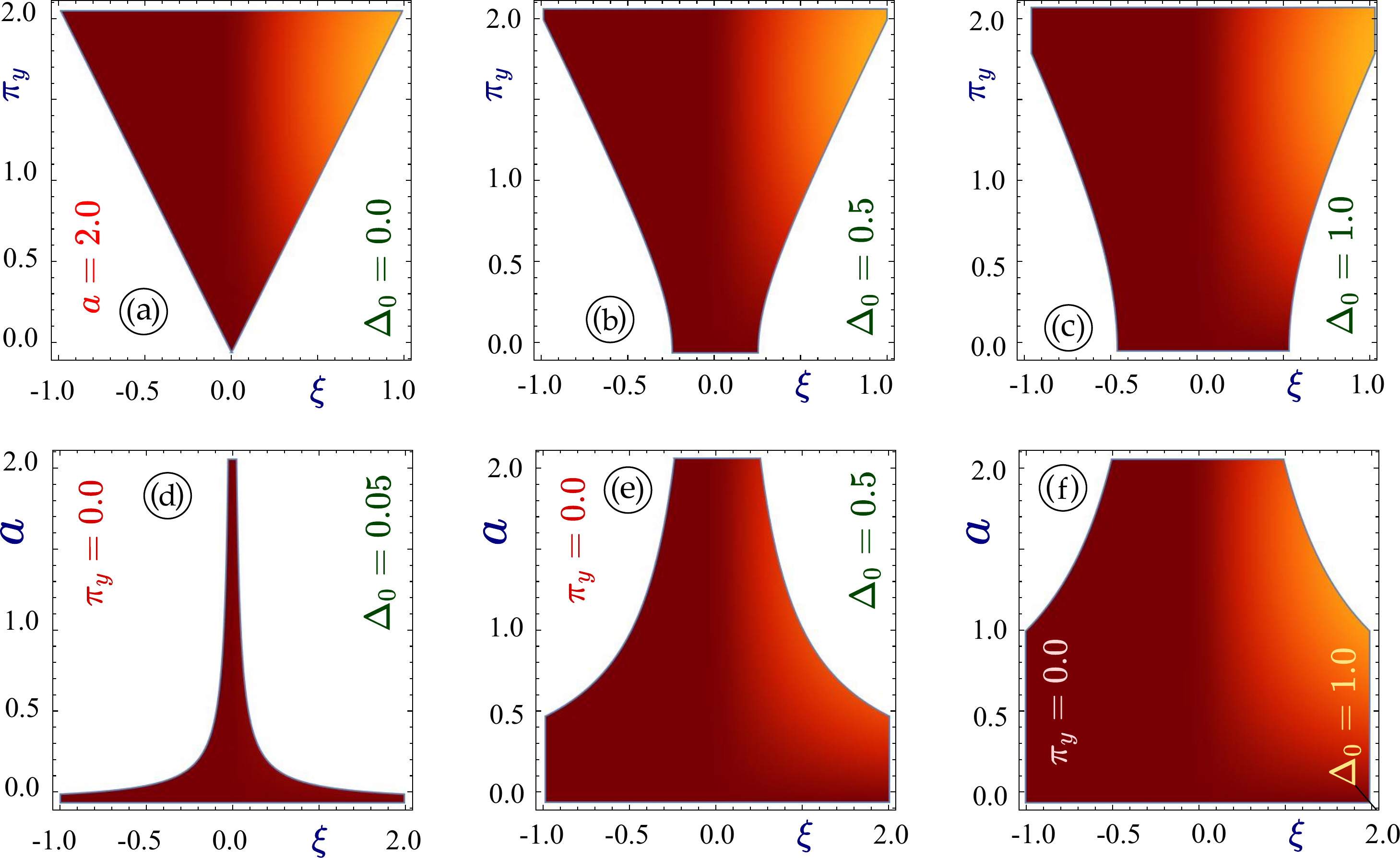}
\caption{(Color online) Plots for classically inaccessible regions, i.e., ${\rm Im}[\pi_x(\xi,\Delta_0\,\vert\,\phi)]\neq 0$, of an incident electron in gapped graphene under a potential barrier $\nu(\xi) = \nu_0 + a \, \xi$. Here, $\varepsilon$ is the kinetic energy of incoming electron,  and $\xi=0$ is selected as the crossing point at which the particle switches between electron and hole states in the barrier region. The boundaries of shaded region correspond to the turning points satisfying $\pi_x(\xi,\Delta_0) = 0$, and  each panel shows the shaded region for specific $\Delta_0$ values as indicated. The upper panels $(a)$--$(c)$ demonstrate how the area and shape of shaded regions depend on $\pi_y$, while the lower panels $(d)$--$(f)$ on the barrier slope $a$. All quantities and parameters used in this graphs and others later are dimensionless and scaled by Eq.\,\eqref{units}.}
\label{FIG:3}
\end{figure}

\begin{figure} 
\centering
\includegraphics[width=0.65\textwidth]{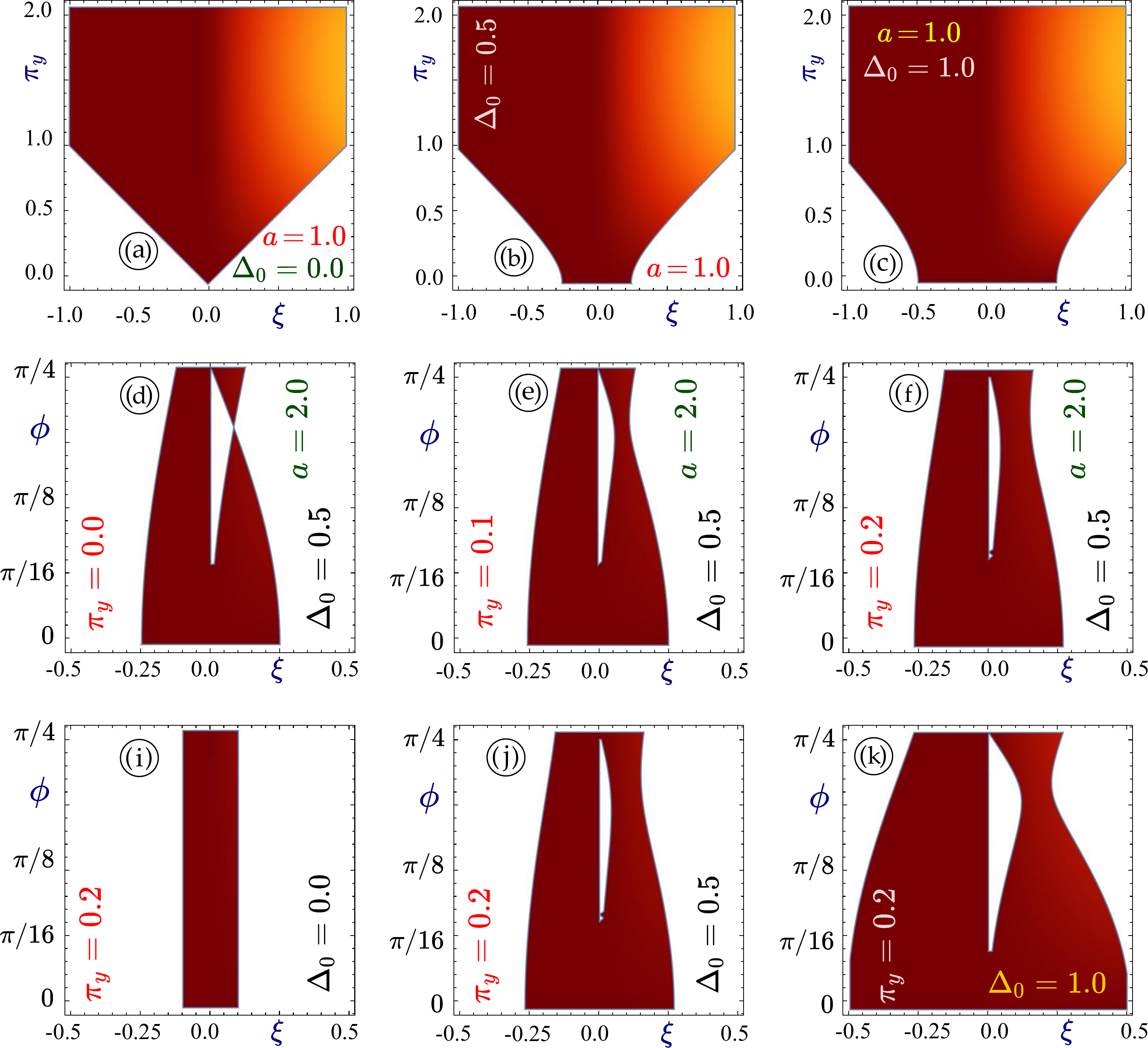}
\caption{(Color online) Plots for classically inaccessible regions (shown shaded) of an incident electron in $\alpha-\mc{T}_3$ materials with $\nu(\xi) = \nu_0 + a \, \xi$.  The boundaries of shaded region correspond to the turning points satisfying $\pi_x(\xi,\Delta_0\,\vert\,\phi) = 0$, and each panel shows the shaded region for specific $\Delta_0$ and $\pi_y$ values as indicated.  Three upper panels show the results as  functions o $\pi_y$ for a dice lattice with $\phi = \pi/4$ and $a=1$,  while the remaining six plots for $\alpha-\mc{T}_3$ lattices as  functions of $\phi$ for various values of $\Delta_0$ and $a=2$.}
\label{FIG:5}
\end{figure}

\begin{figure} 
\centering
\includegraphics[width=0.5\textwidth]{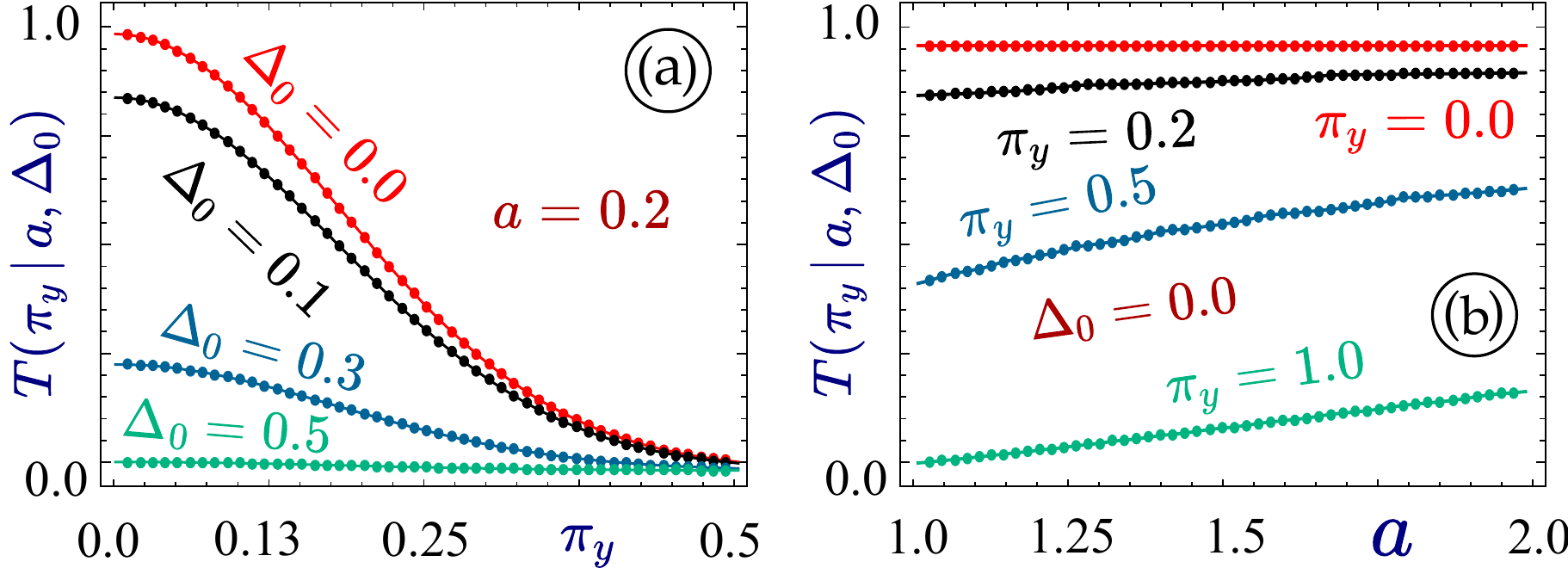}
\caption{(Color online) Calculated transmissions $T\left( \pi_y \, \vert \, a, \Delta_0 \right)$ from Eq.\,\eqref{Tev} for gapped graphene.  Panel $(a)$ presents $T\left( \pi_y \, \vert \, a, \Delta_0 \right)$ as a function of transverse momentum $\pi_y$ for slope $a=0.2$ and various energy gaps $\Delta_0 = 0,\, 0.1,\, 0.3,\, 0.5$.  Plot $(b)$ shows the $a$ dependence of $T\left( \pi_y \, \vert \, a, \Delta_0 \right)$ with $\Delta_0 = 0$ and various vales of $\pi_y = 0,\, 0.2,\, 0.5,\, 1$.}
\label{FIG:4}
\end{figure}

\begin{figure} 
\centering
\includegraphics[width=0.75\textwidth]{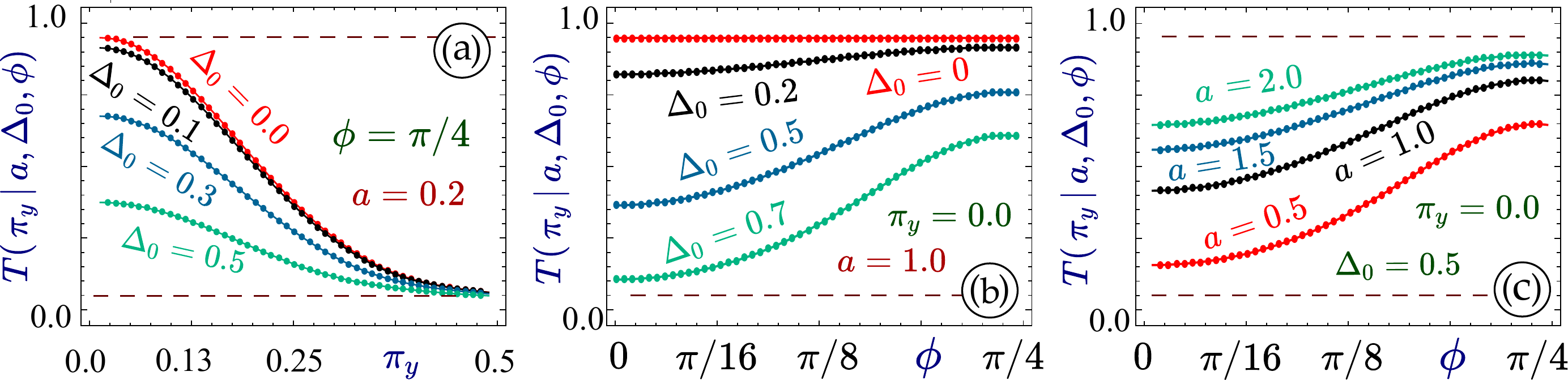}
\caption{(Color online) Calculated transmissions $T\left( \pi_y \, \vert \, a, \Delta_0,\phi \right)$ from Eq.\,\eqref{Tev} for gapped $\alpha-\mc{T}_3$ lattices. 
Panel $(a)$ demonstrates $T\left( \pi_y \, \vert \, a, \Delta_0,\phi \right)$ as a function of transverse momentum $\pi_y$ for $\phi=\pi/4$, $a=0.2$ 
and various energy gaps $\Delta_0 = 0,\, 0.1,\, 0.3,\, 0.5$.  Plots $(b)$ and $(c)$ display the $\phi$ dependence of $T\left( \pi_y \, \vert \, a, \Delta_0,\phi \right)$ with $\pi_y=0$, $a=1$ and various vales of $\Delta_0 = 0,\, 0.2,\, 0.5,\, 0.7$ in $(b)$ 
and with $\pi_y=0$, $\Delta_0=0.5$ and different values of $a=0.5,\,1,\,1.5,\,2$ in $(c)$.}
\label{FIG:6}
\end{figure}

The solution to Eq.\,\eqref{mainEq2} is easily found to be

\begin{equation}
\label{genint}
\varphi_H^{(0)} (\xi) = c_0 \, \tet{exp} \left[
- \int\limits_{\xi_0}^{\xi} \frac{\mc{F}(\zeta \, \vert \, \phi, \tau)}{\Gamma_{(+)}(\zeta \, \vert \,\tau)}\, d\zeta \right] \ .
\end{equation}
Specifically, for a dice lattice with $\phi = \pi/4$, Eqs.\,\eqref{mainEq2} and \eqref{genint} are reduced to 

\begin{equation}
\frac{d\varphi_H^{(0)}(\xi)}{\varphi_H^{(0)}(\xi)} = - \frac{1}{2\,\Gamma_{(+)}(\xi \, \vert \,\tau)} \,
\frac{\pr \, \Gamma_{(+)}(\xi \, \vert \,\tau)}{\pr \xi} \, d \xi = - \frac{\pi_y^2}{\sqrt{\pi_x^2(\xi)-\pi_y^2}} \,\, d\pi_x(\xi) \ ,
\label{new-5}
\end{equation}
while for gapless graphene with $\phi = 0$ we acquire the same results as in Ref.\,[\onlinecite{zalip}].  At last, Eq.\,\eqref{new-5} leads to the solution 

\begin{equation}
\label{vph}
\varphi_H^{(0)}(\xi) = \sqrt{\frac{\pi^2_x(\xi)-\pi_y^2}{\pi^2_x(\xi)}} = \sqrt{1 - \left[\frac{\pi_y}{\pi_x(\xi)}\right]^2 \,} \ ,
\end{equation}
and then $\Psi_0(\xi\,\vert\,\phi,\tau)$ in Eq.\,\eqref{wf0theta} can be completely determined. All other higher-order wave function $\Psi_\lambda(\xi\,\vert\,\phi,\tau)$  for $\lambda=1,\,2,\,\cdots$ can be found by using Eq.\,\eqref{transeq} repeatedly. Moreover, Eq.\,\eqref{vph} becomes divergent as $\pi_x(\xi) = 0$, which is similar to the resulst of Schr\"odinger particles and graphene with a zero or finite bandgap.\,\cite{zalip}  Such a unique feature indicates that the WKB approximation cannot be used in the vicinity of so-called turning points with $\pi_x(\xi) = 0$.

\section{Energy Bandgap in a Pseudospin-1 Lattice}
\label{sec3} 

In this section, we will concentrate on calculating electron transmission over a barrier with a linear potential profile $\nu(\xi) = \nu_0 + 
a \, \xi$ for an $\alpha-\mc{T}_3$ material, where the constant $a$ quantifies the strength of an applied electric field. In fact, we are able to find the transmission solely based on the semi-classical action $\mbb{S}(\xi)$ and the longitudinal component of electron momentum $\pi_x(\xi)$, and we do not need compute  the wave function, its phase factors or the spatial dependence. This is obviously an advantage of employing the WKB approximation, i.e.,  a possibility to acquire very precise transmission of electrons with limited knowledge on electronic states, and an easily evaluated $\mbb{S}(\xi)$  even for a very complicated model Hamiltonian.  In this paper, we focus on two special cases with a bandgap induced either by adding an insulating substrate to an $\alpha-\mc{T}_3$ layer or by imposing an external off-resonance dressing field.  
\medskip 

We start with adding an $\alpha$-independent energy gap $\Delta_0$ to our previous Hamiltonian in Eq.\,\eqref{H0} by using $\hat{\Sigma}_z^{(3)}$ in Eq.\,\eqref{szdice}, namely

\begin{equation}
\label{h1}
\hat{\mc{H}}_{\alpha} (\mbox{\boldmath$k$} \, \vert \, \tau, \phi) = \hbar v_F 
\left[
\begin{array}{ccc}
0 & k^\tau_- \, \cos \phi &  0 \\
k^\tau_+ \, \cos \phi & 0 & k^\tau_- \, \sin \phi \\
0 & k^\tau_+ \, \sin \phi & 0
\end{array}
\right]
+ \Delta_0\left[
\begin{array}{ccc}
1 & 0 & 0 \\
0 & 0 & 0 \\
0 & 0 & -1
\end{array}
\right]\ ,
\end{equation}
which gives rise to an eigenvalue equation

\begin{equation}
\label{ed1}
\varepsilon \left[ \varepsilon^2 - \Delta_0^2 - (\hbar v_F k)^2  \right] + (\hbar v_F k)^2 \Delta_0 \cos2 \phi = 0 \ .
\end{equation}
A similar gap model was adopted in Ref.\,[\onlinecite{Gusgap}] for studying effects of an ionized impurity atom on electronic states of $\alpha-\mc{T}_3$.
\medskip

Figure\ \ref{FIG:1} displays the calculated energy dispersion from Eq.\,\eqref{ed1}, from which we find that symmetry between the valence and conduction bands  under $\Delta_0>0$ is broken for all values of $\phi$ except for $\phi = \pi/4$ (a dice lattice).  Meanwhile, both the bandgap between the valence and flat bands $\Delta_{FV}$ and that between the flat and conduction bands $\Delta_{CF}$ are opened up,  and satisfies $\Delta_{CF}\neq\Delta_{FV}$ away from $k=0$ for all considered cases.  Therefore, the triple connection for the valence, flat and conduction bands in each corner of the Brillouin zone is fully broken.
\medskip 

In the presence of the energy gap $\Delta_0$, the operator $\hat{\mbb{O}}_{\,T} (\xi, \pi_y \, \vert \, \phi, \, \tau)$ initially introduced in Eq.\,\eqref{to1} is modified into 

\begin{equation}
\hat{\mbb{O}}_{\,T} (\xi, \pi_y, \Delta_0 \, \vert \, \phi, \tau) = 
\, \left[
\begin{array}{ccc}
\kappa(\xi) + \Delta_0 & \cos \phi \, \left( \pr\mbb{S}_\Delta(\xi)/\pr\xi - i \pi_y \right) & 0 \\
\cos \phi \, \left( \pr\mbb{S}_\Delta(\xi)/\pr\xi + i \pi_y \right) & \kappa(\xi) & 
\sin \phi \, \left( \pr\mbb{S}_\Delta(\xi)/\pr\xi - i \pi_y \right) \\
0 & \sin \phi \, \left( \pr\mbb{S}_\Delta(\xi)/\pr\xi + i \pi_y \right) & \kappa(\xi) - \Delta_0
\end{array}
\right] \ , 
\label{opt-1}
\end{equation}
where the bandgap $\Delta_0$ should be rescaled to $\Delta_0/V_0$, corresponding to the scaled 
particle's kinetic energy $\varepsilon$ and external potential $\nu(\xi)$ in Eq.\,\eqref{units}. For simplicity, however, we will still adopt the same notation $\Delta_0$.
Here, we would like to emphasize that the Hamiltonian in Eq.\,\eqref{h1} provides a good description for all electronic properties of $\alpha-\mc{T}_3$ lattices, including a dice lattice. 
However, it suffers from a serious limitation in the case of $\phi \rightarrow 0$ for the graphene since the gap term involving $\hat{\Sigma}_z^{(3)}$ in Eq.\,\eqref{h1}
cannot be properly transformed to a $2 \times 2$ Pauli matrix. Therefore, the proper limit for graphene could only be obtained in the limit of $\Delta_0=0$.   
\medskip

To overcome these limitations, we introduce an alternative model Hamiltonian, i.e., 
including a $\phi$-dependent term 

\begin{equation}
\label{h2}
\hat{\mc{H}}_{\Delta} (\phi) = \frac{\Delta}{2} \, \hat{S}_z(\phi) \, 
= \Delta \, \left[
\begin{array}{ccc}
\cos^2 \phi & 0 & 0 \\
0 & - \cos2 \phi & 0 \\
0 & 0 & - \sin^2 \phi
\end{array}
\right] \ ,
\end{equation}
where the bandgap $\Delta$ is included through  

\begin{equation}
\hat{S}_z(\phi) = -i \,
\left[\hat{S}_x(\phi), \hat{S}_y(\phi) 
\right] \ ,
\end{equation}
as employed in Ref.\,[\onlinecite{tutul1}]. Here, the $\phi$-dependent gap term in Eq.\,\eqref{h2} can also be viewed as a part of the Floquet-Magnus Hamiltonian for electron dressed state under a circularly-polarized dressing field,\,\cite{ourpeculiar} which also depends on the valley index $\tau = \pm 1$. It is easy to show that Eq.\,\eqref{h2}, in the graphene limit $\phi\to 0$, reduces to 

\begin{equation}
\hat{\mc{H}}_{\Delta} (\phi \rightarrow 0)= \Delta \left[
\begin{array}{cc|c}
1 & 0 & 0 \\
0 & -1 & 0 \\
\hline
0 & 0 & 0
\end{array}
\right]\ ,
\end{equation}
and meanwhile, for a dice lattice with $\phi = \pi/4$, to 

\begin{equation}
\hat{\mc{H}}_{\Delta} (\phi \rightarrow \pi/4) = \frac{\Delta}{2}
\left[
\begin{array}{ccc}
	1 & 0 & 0 \\
	0 & 0 & 0 \\
	0 & 0 & -1
\end{array}
\right]=\frac{\Delta}{2}\,\hat{\Sigma}_z^{(3)}\ , 
\end{equation}
which implies that the effect from laser irradiation on a dice lattice is only half of that on graphene. 
\medskip

By combining Eqs.\,\eqref{H0} and \eqref{h2}, the energy dispersion for gapped $\alpha-\mc{T}_3$ lattices is found to satisfy the following equation, i.e.,

\begin{equation}
\label{ed2}
\varepsilon^3 - \left(
\frac{5 \Delta^2}{8} + k^2
\right)\varepsilon - \frac{\Delta^2}{8} \left[
3 \,\varepsilon \, \cos 4 \phi + \frac{\Delta}{3} \, \sin 2 \phi \,  \sin 4 \phi\right] = 0 \ ,
\end{equation}
which gives rise to three solutions, given by

\begin{equation}
\label{as1}
\varepsilon_\lambda(k,\Delta\,\vert\,\phi)= \frac{2}{\sqrt{3}} \, \sqrt{
k^2 + \frac{\Delta^2}{8} \left(
5 + 3 \cos4 \phi
\right) \,}\,
\cos \left[ \frac{2 \pi \lambda}{3} + 
\cos^{-1} \left( \frac{9 \sqrt{6}\, \Delta^3 \,\sin2 \phi \, \sin4 \phi}{\left[
8 k^2 + \Delta^2(5 + 3 \cos4 \phi)
\right]^{3/2}
} \right) 
\right] \ , 
\end{equation}
where $\lambda = 0,\,1,\,2$ specifies three different energy bands.
\medskip

The calculate energy dispersions from Eq.\,\eqref{as1} are presented in Fig.\,\ref{FIG:2}, and the results for both graphene ($\phi=0$) and dice lattice ($\phi=\pi/4$) are displayed  in the top and bottom rows, respectively, along with the case with $\phi=\pi/6$ for general $\alpha-\mc{T}_3$ material in the middle row.  From both top and bottom rows, we find symmetric dispersions with respect to $k=0$, and the graphene gap at $k=0$ is exactly twice of that for a dice lattice. As $\phi=\pi/6$ for the middle row, the middle ``flat band'' becomes dispersive in the way opposite to that in panels $(a)$--$(c)$ of Fig.\,\ref{FIG:1}, i.e., switching between peak and valley  at $k=0$. Moreover, the mirror symmetry between the valence and conduction band is also broken for a finite value of $\Delta$ and all values of $0<\phi<\pi/4$.

\medskip
\par

On the other hand, after a finite bandgap has been taken into account by Eq.\,\eqref{transeq}, the previous operator $\hat{\mbb{O}}_{\,T} (\xi, \pi_y\, \vert \, \phi, \tau)$ 
in Eq.\,\eqref{opt-1} has been changed to 

\begin{equation}
\hat{\mbb{O}}_{\,T} (\xi, \pi_y, \Delta_0 \, \vert \, \phi, \tau) = 
\, \left[
\begin{array}{ccc}
\kappa(\xi) + \Delta_0 \cos^2 \phi & \cos \phi \, \left(\pr\mbb{S}_\Delta(\xi)/\pr\xi - i \pi_y \right) & 0 \\
\cos \phi\left(\pr\mbb{S}_\Delta(\xi)/\pr\xi + i \pi_y \right) & \kappa(\xi) - \Delta_0 \cos2\phi & 
\sin \phi\left(\pr\mbb{S}_\Delta(\xi)/\pr\xi - i \pi_y \right) \\
0 & \sin \phi\left(\pr\mbb{S}_\Delta(\xi)/\pr\xi + i \pi_y \right) & \kappa(\xi) - \Delta_0 \sin^2 \phi
\end{array}
\right] \ , 
\label{opt-2}
\end{equation}
where $\Delta_0=\Delta/V_0$. Therefore, from Eqs.\,\eqref{opt-2} and \eqref{sp} we are able to find explicitly 
the spatially-dependent longitudinal momentum $\pi_x(\xi, \Delta_0\,\vert\,\phi)$ as 

\begin{equation}
\label{pixd2}
[\pi_x(\xi, \Delta_0 \, \vert \, \phi)]^2 = \kappa^2(\xi) - \pi_y^2 - \frac{\Delta_0^2}{8} \left(5 + 3 \cos4 \phi\right)+ 
\frac{\Delta_0^2}{8 \kappa(\xi)} \, \sin2 \phi \, \sin4 \phi \ . 
\end{equation}
As $\phi=0$, w get from Eq.\,\eqref{pixd2} that $[\pi_x(\xi, \Delta_0 \,\vert\, \phi = 0)]^2 = \kappa^2(\xi) - \Delta_0^2 - \pi_y^2$, which is the same as that in Ref.\,[\onlinecite{zalip}]. 
For a dice lattice with $\phi=\pi/4$, on the other hand, we find $[\pi_x(\xi, \Delta_0 \,\vert\, \phi = \pi/4)]^2 = \kappa^2(\xi) - (\Delta_0/2)^2 - \pi_y^2$. 

\medskip
\par

Equation\ \eqref{pixd2} becomes quadratic if its last term equals zero, which can be satisfied for either graphene with $\phi = 0$ or a dice lattice with $\phi = \pi/4$.  In these two case, the classically inaccessible regions are simply connected, as seen in Figs.\,\ref{FIG:3} and \ref{FIG:5}.  For all other $\phi$ values, the turning points, or the boundaries of classically forbidden regions, are determined by a cubic $\kappa(\xi)$ equation and  these regions consist of several parts with non-trivial shapes and connections, as demonstrated in Fig.\,\ref{FIG:5}$(d)$--$(k)$.  

\medskip
\par

In WKB theory, the transmission amplitude $T(\pi_y \, \vert \, a, \Delta_0, \phi)$,  or the probability for electron tunneling, can be estimated by the integral of $|\pi_x(\xi,\Delta_0\,\vert\,\phi)|$ presented in Eq.\,\eqref{pixd2},  which is equivalent to ${\rm Im}[\pi_x(\xi,\Delta_0\,\vert\,\phi)]$, over the classically forbidden regions (CFR). This leads to \,\cite{sonin, fanwar}   

\begin{equation}
\label{Tev}
T(\pi_y \, \vert \, a, \Delta_0, \phi) = \tet{exp} \left[ 
- \frac{2}{\hbar_0} \, \int\limits_{\rm CFR} \vert \pi_x(\xi) \vert \, d \xi 
\right]
\end{equation}
under the condition of

\begin{equation}
\int\limits_{\rm CFR} \vert \pi_x(\xi) \vert \, d \xi  \gg \hbar_0 \ . 
\end{equation}
Here, CFR are defined by $\pi^2_x(\xi, \Delta_0)<0$, i.e., the particle acquires an imaginary longitudinal momentum as then a strongly decayed transmission.  The calculated location and size of CFR from Eq.\,\eqref{pixd2} for gapped graphene are presented as shaded regions in Fig.\,\ref{FIG:3}.
The boundaries of CFR, determined by $\pi_x(\xi, \Delta_0) = 0$, are not linear for $\kappa(\xi)$ in the presence of a finite energy gap, as seen from Eq.\,\eqref{pixd2}.  The obtained CFR is always symmetric with the selected electron-to-hole crossing point $\xi = 0$, given by $\kappa(\xi)= [\nu(\xi) - \varepsilon]/\hbar_0 = 0$.  For the case of $\Delta_0 = 0$, we find from Eq.\,\eqref{pixd2} that $\kappa(\xi) = -\sqrt{\pi_x^2(\xi)+\pi_y^2}\equiv -\pi_{tot}(\xi)$  which is opposite to the total momentum $\pi_{tot}(\xi)$.

\medskip
\par

Even though the result in Eq.\,\eqref{Tev} appears only as an estimation, for some trivial cases, e.g., gapless graphene, this result becomes accurate.\,\cite{sonin}  In fact, for $\Delta_0=0$ we know that both Eq.\,\eqref{pixd2} for $\pi_x(\xi)$ and the conditions for determining the turning points $\pi_x(\xi) = 0$  coincide with gapless graphene. Therefore, the obtained transmission $T(\pi_y\,\vert\, a)=\tet{exp}[-(\pi/a)\,\pi_y^2]$ is expected 
also true for an arbitrary $\alpha-\mc{T}_3$ material with $0 < \alpha < 1$ as long as $\Delta_0 = 0$.   Furthermore, we expect that the width of CFR will increase with $\pi_y$, as verified from Eq.\,\eqref{pixd2},   since it leads to decrease of $\pi_x(\xi)$ for a given energy $\varepsilon$ of incoming particle.  However, the CFR width reduces for increasing slope $a$ because the electron/hole crossover and the new classically allowable state is now achieved within a shorter displacement  of a charger particle along its trajectory under a non-uniform potential.  These predicted effects are indeed observed for electron transmission in gapped graphene, as presented in Fig.\,\ref{FIG:4}.  Apart from the energy gap $\Delta_0$, which always results in reduced transmission, we see a similar effect from $\pi_y$ kept as a constant in our model, as seen in Fig.\,\ref{FIG:4}$(a)$.  In contrast, a larger potential slope $a$ leads to an enhanced transmission for fixed $\Delta_0$, as found from Fig.\,\ref{FIG:4}$(b)$.

\medskip
\par

Results for transmission of electrons in gapped $\alpha-\mc{T}_3$ are presented in Fig.\,\ref{FIG:6}.  Similar to gapped graphene, the Klein paradox occurs only for a head-on collision, or $\pi_y  = 0$, and $\Delta_0=0$, as found from Figs.\,\ref{FIG:6}$(a)$ and \ref{FIG:6}$(b)$.Moreover, we also find that the Klein paradox remains true for all values of $a$ and $\phi$, as long as these two conditions are met.  For a finite bandgap $\Delta_0 \geq 0.2$ for $\alpha-\mc{T}_3$, we see a strong dependence of transmission on phase $\phi$,  which becomes considerable less than one as shown in Figs.\,\ref{FIG:6}$(b)$ and \ref{FIG:6}$(c)$.  In this case, transmission increases with $\phi$ due to existence of the flat band in energy dispersion. 
Furthermore, we also find increasing transmission with $a$ as seen in Fig.\,\ref{FIG:6}$(c)$. Such a feature could be attributed to the fact that switching between electron and hole states will be faster for a steeper potential, and this explanation 
holds true for both gapped graphene and all types of $\alpha-\mc{T}_3$ materials.

\section{Concluding Remarks and Summary}
\label{sec4}

In conclusion, we have generalized the WKB semi-classical approximation for pseudospin-1 $\alpha-\mc{T}_3$ lattices by deriving a complete set of recurrence transport equations.  The solutions of these coupled differential equations have provided the semi-classical wave functions for gapped and the phase dependent $\alpha-\mc{T}_3$ Hamiltonian,  and led to the correct description of quantum states for charge carriers in the ballistic regime.  Additionally, we have   obtained closed form analytic expression for WKB wave functions which could be applied to various analytical models as well as for studying  the tunneling properties of electrons in gapped $\alpha-\mc{T}_3$ materials.

\medskip
\par

Our derivation of the generalized WKB equations and a pseudospin-1 Dirac-Weyl Hamiltonian for the $\alpha-\mc{T}_3$ model are shown to be quite different from a Schr{\"o}dinger particle considered in standard quantum mechanics and even a Dirac electron in graphene. For this case, we are facing with $3 \times 3$ phase dependent matrices and inequivalent states of electrons and holes. In particular, the resulting electronic states and even their energies are found to be valley-dependent under an external radiation field,  
which enables the so-called valley filtering effect and a plenty of other valleytronic applications.

\medskip
\par

Physically, our derived dynamical equations and calculated electronic states could be employed for studying a variety of transport properties of $\alpha-\mc{T}_3$ materials. However, our focus in this paper was on the phase dependent electron tunneling, band gap modification and the suppression of the Klein paradox with non-square potential barriers. Even a limited knowledge regarding the semi-classical action and time-dependent momentum can provide us with important information on electron dynamics and help us evaluate the  tunneling transmission of electrons through the integral of absolute electron momentum over the classically inaccessible regions along the tunneling-electron path.

\medskip
\par

Based on our generalized WKB theory, we have investigated the electron tunneling through an electric field biased potential barrier and revealed unimpeded Klein tunneling for the head-on collisions in the absence of an energy gap. This event applies to all $\alpha-\mc{T}_3$ materials independent of the geometry phase $\phi$,  i.e., the calculated transmission in gapless $\alpha-\mc{T}_3$ materials does not depend on $\phi$. Moreover, we have found that the $\pi_y$-dependent transmission is greatly reduced  in the presence of a band gap, and is  decreased when $\pi_y$ is increased. On the other hand, the slope of the potential profile always enhances the transmission despite the energy gap,  which is due to speeding up an electron-to-hole (or hole-to-electron) transition over a shortened distance in $\xi$ position space. As a result, the generalized WKB theory in this paper could be utilized to discern the array of localized and trapped electronic states through their barrier scattering effects.

\medskip
\par

Our effort in generalizing the WKB approximation to deal with the Hamiltonian for $\alpha-\mc{T}_3$ lattices provides additional tools to explore additional important and unknown modifications to  tunneling mechanism in such materials, as demonstrated by our obtained analytical expressions for electron transmission and Klein tunneling.  We believe that our current study has revealed the most remarkable and exclusive physics features of novel low-dimensional materials.  Meanwhile, all these discoveries will definitely find their applications in Dirac cone based tunneling transistors in analog RF devices, along with their tunneling current control  by constructive barrier scattering across designed array of coherent scatters. 

\begin{acknowledgements}
A.I. would like to acknowledge the funding received from TRADA-51-82 PSC-CUNY Award $\#$ 63061-00-51. D.H. was supported by the Air Force Office of Scientific Research (AFOSR).  G.G. would like to acknowledge Grant No. FA9453-21-1-0046 from the Air Force Research Laboratory (AFRL).
\end{acknowledgements}

\bibliography{WKBbib}

\end{document}